\documentclass{article}

     \PassOptionsToPackage{numbers, compress}{natbib}
\usepackage[preprint]{neurips_2025}

\usepackage[utf8]{inputenc} % allow utf-8 input
\usepackage[T1]{fontenc}    % use 8-bit T1 fonts
\usepackage{hyperref}       % hyperlinks
\usepackage{url}            % simple URL typesetting
\usepackage{booktabs}       % professional-quality tables
\usepackage{amsfonts}       % blackboard math symbols
\usepackage{nicefrac}       % compact symbols for 1/2, etc.
\usepackage{microtype}      % microtypography
\usepackage{xcolor}         % colors
\usepackage{amsmath}
\usepackage{graphicx} 
\usepackage{wrapfig}

\usepackage{xcolor}

\title{Surf2CT: Cascaded 3D Flow Matching Models for Torso 3D CT Synthesis from Skin Surface}

\author{%
  Siyeop Yoon\textsuperscript{1}\thanks{Email: \texttt{syoon5@mgh.harvard.edu, yooneige@gmail.com}}
  \quad
  Yujin Oh\textsuperscript{1}
  \quad
  Pengfei Jin\textsuperscript{1}
  \quad
  Sifan Song\textsuperscript{1}
  \quad
  Matthew Tivnan\textsuperscript{1}\\
  \quad
  \textbf{Dufan Wu\textsuperscript{1}}
  \quad
  \textbf{Xiang Li\textsuperscript{1}}
  \quad
  \textbf{Quanzheng Li\textsuperscript{1}}\thanks{Email: \texttt{li.quanzheng@mgh.harvard.edu}}\\
  \textsuperscript{1} Massachusetts General Hospital and Harvard Medical School, Boston, MA 02114
}

\begin{document}

\maketitle

\begin{abstract}
We present \emph{Surf2CT}, a novel cascaded \emph{flow matching} framework that synthesizes full 3D computed tomography (CT) volumes of the human torso from external surface scans and simple demographic data (age, sex, height, weight). This is the first approach capable of generating realistic volumetric internal anatomy images solely based on external body shape and demographics, without any internal imaging. Surf2CT proceeds through three sequential stages: (1) \textbf{Surface Completion}, reconstructing a complete signed distance function (SDF) from partial torso scans using conditional 3D flow matching; (2) \textbf{Coarse CT Synthesis}, generating a low-resolution CT volume from the completed SDF and demographic information; and (3) \textbf{CT Super-Resolution}, refining the coarse volume into a high-resolution CT via a patch-wise conditional flow model. Each stage utilizes a 3D-adapted EDM2 backbone trained via flow matching. We trained our model on a combined dataset of 3,198 torso CT scans (approximately 1.13 million axial slices) sourced from Massachusetts General Hospital (MGH) and the AutoPET challenge. Evaluation on 700 paired torso surface-CT cases demonstrated strong anatomical fidelity: organ volumes exhibited small mean percentage differences (range from -11.1\% to 4.4\%), and muscle/fat body composition metrics matched ground truth with strong correlation ($R^2$ has range from 0.67 to 0.96). Lung localization had minimal bias (mean difference -2.5 mm), and surface completion significantly improved metrics (Chamfer distance: from 521.8 mm to 2.7 mm; Intersection-over-Union: from 0.87 to 0.98). Surf2CT establishes a new paradigm for non-invasive internal anatomical imaging using only external data, opening opportunities for home-based healthcare, preventive medicine, and personalized clinical assessments without the risks associated with conventional imaging techniques.
\end{abstract}
\section{Introduction}
Reconstructing internal anatomical structures without medical imaging equipment remains a significant challenge in biomedical imaging. Computed tomography (CT) offers detailed volumetric information but involves high costs and radiation exposure, making routine use and large-scale screening impractical \cite{smith2009radiation}. Generating clinically informative 3D CT volumes from non-invasive external body shape and basic demographic data (age, sex, height, and weight) presents an attractive alternative. However, external body shape alone does not uniquely determine internal anatomy, as similar external measurements can correspond to diverse internal organ sizes and compositions \cite{treleaven20073d, lee2021deep}.

Addressing this challenge, we introduce \textbf{Surf2CT}, a cascaded generative framework synthesizing realistic internal 3D CT images directly from external body surfaces and demographic information. The clinical implications of this capability are profound. For instance, preventive medicine could leverage inexpensive 3D body scanners in primary care or even home settings to non-invasively capture external shapes. Subsequently, our model would generate personalized CT-like volumetric data, enabling clinicians to assess health parameters such as organ dimensions, visceral fat accumulation, or muscle mass without ionizing radiation exposure \cite{despres2008abdominal,cruz2019sarcopenia}. Moreover, in surgical planning and simulation contexts, approximate internal anatomical models could serve as radiation-free digital twins, aiding surgeons through augmented reality training and preoperative preparations \cite{asciak2025digital}. 

Recent advances in generative modeling, notably diffusion models and flow-matching methods, have successfully synthesized images and shapes from sparse or ambiguous inputs \cite{song2019generative, lipman2022flow, vahdat2022lion, galvis2024sc}. However, these have yet to tackle the task of inferring internal volumetric anatomy directly from external surfaces and demographics, as existing approaches typically depend on detailed internal segmentations or sparse imaging data \cite{guo2025maisi, shen2019patient, pan2024synthetic}.

\textbf{Surf2CT} addresses these limitations through a three-stage approach: (1) reconstructing a complete torso geometry from incomplete surface scans using a signed distance function (SDF), (2) generating a coarse-resolution CT volume conditioned on the reconstructed SDF and demographic data, and (3) refining this volume through super-resolution, enhancing organ boundaries and tissue interfaces for anatomical consistency. The key contributions of this work are:
\begin{itemize}
\item Introducing \textbf{Surf2CT}, the first framework capable of synthesizing clinically relevant, high-fidelity 3D CT volumes solely from external surfaces and demographic data.
\item Employing a novel three-stage pipeline—surface completion, conditional CT generation, and anatomical refinement—to effectively address the surface-to-volume inference challenge.
\item Demonstrating Surf2CT’s robustness and clinical applicability through rigorous evaluation across organ volume accuracy, body composition (muscle and fat), lung localization, and surface reconstruction quality.
\end{itemize}

\section{Related Work}
\paragraph{Statistical Shape Models and Internal Anatomy}
Digital anthropomorphic phantoms, such as the XCAT family \cite{segars20104d}, provide detailed 3D models of internal organs and tissues for simulating medical imaging. XCAT enables realistic CT and PET simulations by parameterizing organ shape, physiology, and motion, but relies on a limited template library with only coarse demographic scaling and manual adjustments for individual variability. Additionally, XCAT models are not easily modifiable and struggle to represent the full diversity of patient-specific anatomy.

Recent work like BOSS \cite{shetty2023boss} learns a joint statistical model of skin, bone, and organ geometry directly from CT data, producing a deformable digital twin via mesh representation. While BOSS improves anatomical modeling, it outputs only surface geometry (e.g., point clouds or meshes) and not volumetric intensity data. These models help bridge external and internal anatomy, but none synthesize patient-specific CT volumes directly from external scans and demographic data.

\paragraph{CT Reconstruction from Sparse X-ray Views}
A separate line of work addresses 3D CT reconstruction from sparse or limited-view radiographs. Ying et al.\ proposed X2CT-GAN, a 3D GAN that reconstructs chest CT from two orthogonal X-ray views \cite{ying2019x2ct}. DIFR3CT \cite{sun2024difr3ct} extends this by learning a conditional 3D diffusion model capable of generating plausible CT volumes from fewer than 10 X-rays. Other approaches similarly rely on calibrated, spatially aligned projections to infer internal structures \cite{shen2019patient,shen2022geometry}. These methods require at least partial internal imaging. In contrast, Surf2CT removes this dependency entirely, synthesizing CT volumes purely from external scans and demographic data.
\paragraph{Segmentation-Guided CT Synthesis}
CT synthesis guided by internal segmentation maps has shown strong performance by imposing anatomical constraints on generative models. Seg2Med \cite{yang2025seg2med} uses organ-level masks to condition image generation, improving realism and anatomical fidelity across multiple modalities. MAISI \cite{guo2025maisi} further applies diffusion models guided by multi-organ masks to synthesize high-resolution CT scans aligned with internal structures. However, both approaches assume access to accurate internal segmentations—typically derived from pre-existing CT or MRI scans. In contrast, Surf2CT synthesizes anatomically accurate CT volumes directly from external body shapes and demographics, eliminating the need for internal segmentation masks and expanding applicability to scenarios lacking prior medical imaging.

\begin{figure}
    \centering
    \includegraphics[width=1\linewidth]{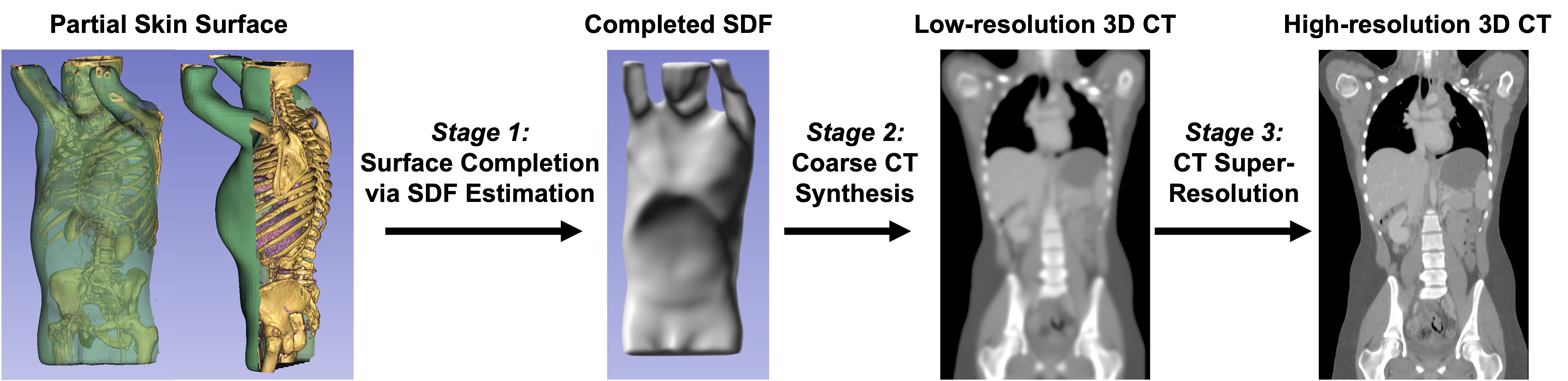}
    \caption{Overview of the proposed cascaded generative pipeline. Our framework consists of three sequential stages: (1) Surface completion using conditional flow matching to estimate a full signed distance function (SDF) from partial scans; (2) coarse CT volume synthesis (at 8 mm isotropic resolution) conditioned on the completed SDF and demographic attributes; and (3) patch-wise CT super-resolution to refine the coarse CT into a detailed high-resolution volume (2 mm isotropic). }
    
    \label{fig:pipeline}
\end{figure}

\section{Method}
We propose \textbf{Surf2CT}, a cascaded generative framework synthesizing anatomically realistic 3D CT volumes from partial surface scans and demographic information (age, binary sex, height, weight). Our approach employs a three-stage conditional flow matching framework, progressively refining Gaussian noise into detailed anatomical structures (Fig.~\ref{fig:pipeline}): (1) Surface completion via signed distance function (SDF) estimation; (2) Coarse CT synthesis; and (3) Patch-wise high-resolution CT refinement. At inference time, the approach conveniently requires only an incomplete external surface scan and the individual's demographic data, without any internal imaging.

\paragraph{Stage 1: Surface Completion via SDF Estimation}

The first stage aims to reconstruct a complete 3D torso shape from a partial surface scan. We represent body geometry using a signed distance function (SDF) \cite{chu2023diffcomplete}, which assigns to each point in space the shortest distance to the body surface, with the sign indicating whether the point lies inside or outside the body. Formally, the SDF is defined as:
\begin{equation}
f(\mathbf{x}) =
\begin{cases}
\phantom{-}\min\limits_{\mathbf{y} \in \partial \Omega} \|\mathbf{x} - \mathbf{y}\|_2, & \text{if } \mathbf{x} \in \Omega, \\
-\min\limits_{\mathbf{y} \in \partial \Omega} \|\mathbf{x} - \mathbf{y}\|_2, & \text{if } \mathbf{x} \notin \Omega,
\end{cases}
\label{eq:sdf}
\end{equation}
where \( \Omega \subset \mathbb{R}^3 \) denotes the volume enclosed by the surface \( \partial \Omega \).  We discretize \( f(\mathbf{x}) \) on a \( 56 \times 56 \times 88 \) with $8\,$mm isotropic grid covering the torso region. From the completed surface, we construct a partial SDF volume \( f_{\text{partial}} \), computed by selectively removing regions of the full body surface—such as retaining only the frontal view—so that valid signed distance values exist near the observed surface, while the rest of the volume remains undefined or filled with heuristic guesses. 

Our objective is to estimate a realistic, anatomically plausible full SDF \( f_{\text{full}} \) consistent with the partial surface and the subject’s demographics. To achieve this, we train a conditional generative model \( G_1 \) using the flow matching:
\begin{equation}
f_{\text{full}} = G_1(\boldsymbol{\xi}; f_{\text{partial}}, \mathbf{z}_{\text{demo}}), \quad \boldsymbol{\xi} \sim \mathcal{N}(0, I).
\end{equation}

This is implemented via an ordinary differential equation (ODE) defined by a learned velocity field:
\begin{equation}
\frac{d\mathbf{x}(t)}{dt} = v_\theta^{(1)}(t, \mathbf{x}(t), f_{\text{partial}}, \mathbf{z}_{\text{demo}}),
\end{equation}
which transports the initial state \( \mathbf{x}(0) \sim \mathcal{N}(0, I) \) to a realistic shape sample \( \mathbf{x}(1) \sim p(f_{\text{full}} \mid f_{\text{partial}}, \mathbf{z}_{\text{demo}}) \). During training, we sample ground-truth pairs \( (f_{\text{partial}}, f_{\text{full}}^{\text{gt}}) \) and use a linear interpolation path:
\begin{equation}
\mathbf{x}_t = (1 - \alpha(t)) \boldsymbol{\eta} + \alpha(t) f_{\text{full}}^{\text{gt}}, \quad \boldsymbol{\eta} \sim \mathcal{N}(0, I),
\end{equation}
with a schedule \( \alpha(t) \in [0, 1] \). The model is trained to match the target velocity:
\begin{equation}
\mathcal{L}_1 = \mathbb{E}_{t, \boldsymbol{\eta}, f_{\text{full}}^{\text{gt}}} \left[ \left\| v_\theta^{(1)}(t, \mathbf{x}_t) - (f_{\text{full}}^{\text{gt}} - \boldsymbol{\eta}) \right\|^2 \right].
\end{equation}
The demographic attributes (age, binary sex, height, weight) are each inputted as separate channels with homogeneous values repeated across the spatial dimensions of the SDF grid, resulting in four additional channels of the same size as \( f_{\text{partial}} \). These are concatenated with the partial SDF and fed directly into the model as input. At inference time, we provide a new subject’s partial surface in the form of \( f_{\text{partial}} \) along with their demographic vector \( \mathbf{z}_{\text{demo}} \). We sample Gaussian noise \( \boldsymbol{\xi} \sim \mathcal{N}(0, I) \) and solve the learned ODE to generate the complete SDF volume \( f_{\text{full}} \).

\paragraph{Stage 2: Coarse CT Synthesis from the Completed Signed Distance Function}

In the second stage, we generate a coarse-resolution CT volume from the completed torso shape (the full SDF) obtained in Stage~1. Specifically, we define a low-resolution CT, \( X_{\text{low}} \), on an isotropic \(8\,\text{mm}\) voxel grid with dimensions \(56\times56\times88\), covering a physical field-of-view of approximately \(448\times448\times704\,\text{mm}\). We train another conditional generative model \(G_2\), again using the flow matching, to synthesize this coarse CT volume from the completed SDF and demographic information:
\begin{equation}
X_{\text{low}} = G_2(\boldsymbol{\xi}_2;\, f_{\text{full}}, \mathbf{z}_{\text{demo}}), \quad \boldsymbol{\xi}_2 \sim \mathcal{N}(0,I).
\end{equation}

This model defines a probability distribution over plausible CT volumes conditioned on the shape and demographic attributes via a learned ODE governed by the velocity field \(v_\theta^{(2)}\):
\begin{equation}
\frac{d\mathbf{x}(t)}{dt} = v_\theta^{(2)}(t,\mathbf{x}(t), f_{\text{full}}, \mathbf{z}_{\text{demo}}), \quad \mathbf{x}(0)\sim\mathcal{N}(0,I), \quad \mathbf{x}(1)\sim p(X_{\text{low}}\mid f_{\text{full}}, \mathbf{z}_{\text{demo}}).
\end{equation}

During training, we utilize ground-truth pairs \((f_{\text{full}}^{\text{gt}}, X_{\text{low}}^{\text{gt}})\), where \(f_{\text{full}}^{\text{gt}}\) is computed from skin surface of the high-resolution CT scans and \(X_{\text{low}}^{\text{gt}}\) is obtained by downsampling these CT scans to the target 8~mm voxel resolution. We employ a linear interpolation training scheme:
\begin{equation}
X_t = (1-\alpha(t))\,\boldsymbol{\eta} + \alpha(t)\,X_{\text{low}}^{\text{gt}}, \quad \boldsymbol{\eta}\sim\mathcal{N}(0,I),
\end{equation}
with a scheduling function \(\alpha(t)\) smoothly transitioning from noise to data. The training objective is:
\begin{equation}
\mathcal{L}_2 = \mathbb{E}_{t, \boldsymbol{\eta}, X_{\text{low}}^{\text{gt}}}\left[\| v_\theta^{(2)}(t, X_t, f_{\text{full}}^{\text{gt}}, \mathbf{z}_{\text{demo}}) - (X_{\text{low}}^{\text{gt}} - \boldsymbol{\eta}) \|^2\right].
\end{equation}

As \( t \to 1 \), this approach ensures that the synthesized CT volumes converge to realistic internal density distributions consistent with both the body shape captured by the completed SDF and demographic attributes, effectively learning anatomical variations correlated with external body geometry and subject-specific information.

After training, we generate a coarse-resolution CT for a new subject by first obtaining the completed SDF \( f_{\text{full}} \) from Stage~1, then sampling the low-resolution CT volume as:
\begin{equation}
X_{\text{low}} = G_2(\boldsymbol{\xi}; f_{\text{full}}, \mathbf{z}_{\text{demo}}), \quad \boldsymbol{\xi}\sim\mathcal{N}(0,I),
\end{equation}
by numerically integrating the learned ODE from noise (\( t=0 \)) to the data distribution (\( t=1 \)). The resulting coarse CT, \( X_{\text{low}} \), captures the large-scale anatomical structure of the subject. Although limited in resolution and somewhat blurry, it clearly delineates major tissue types such as lung regions (low-density areas), soft tissues, fat distribution, and coarse outlines of internal organs.

Importantly, the model \( G_2 \) implicitly learns correlations between external shape features and internal organ positions, sizes, and tissue densities. Conditioning on demographic attributes \( \mathbf{z}_{\text{demo}} \) allows the model to adjust internal structures according to known physiological trends. For instance, older or higher-BMI individuals typically have increased adipose tissue, leading to lower-density regions in predictable anatomical areas. The model leverages demographic conditioning to incorporate these realistic variations even when the shape information alone may be ambiguous.

\paragraph{Stage 3: CT Super-Resolution (Patch-wise High-Resolution Synthesis)}

The final stage enhances the coarse CT volume from Stage~2 into a high-resolution CT with detailed anatomical structures. The target volume, \( X_{\text{high}} \), is defined on the original \(224\times224\times352\) voxel grid, matching the physical dimensions of the full SDF from Stage~1. This corresponds to approximately \(2\,\text{mm}\) isotropic resolution after upsampling the \(8\,\text{mm}\) coarse CT by a factor of 4.

Directly generating full-resolution CT volumes at this scale is computationally expensive. To manage memory consumption and computational load, we adopt a patch-wise training strategy \cite{wang2023patch}. Specifically, the high-resolution generative model \( G_3 \) synthesizes voxel patches of high-resolution CT conditioned on corresponding regions from the coarse CT, positional encodings, and demographics:
\begin{equation}
P = G_3(\boldsymbol{\xi};\, \widetilde{X}_{\text{low}}, \mathbf{z}_{\text{demo}}, \text{pos}), \quad \boldsymbol{\xi}\sim\mathcal{N}(0,I),
\end{equation}
where \(\widetilde{X}_{\text{low}}\) is the coarse CT upsampled via trilinear interpolation to match the high-resolution CT.

During training, we randomly sample high-resolution ground-truth patches \(P^{\text{gt}}\) (each \(56\times56\times88\) voxels) from full-resolution CT volumes. Each input patch is complemented with fixed sinusoidal positional encodings to inform the model explicitly of its spatial location within the full torso. Demographic features are provided as global conditioning channels, with homogeneous values replicated spatially. Thus, the model learns a localized conditional distribution:
\begin{equation}
p\left(P^{\text{gt}}\mid\widetilde{X}_{\text{low}},\mathbf{z}_{\text{demo}},\text{pos}\right).
\end{equation}

We employ flow matching for training, defining an interpolation path from noise to data:
\begin{equation}
\mathbf{x}_t = (1 - \alpha(t))\,\boldsymbol{\eta} + \alpha(t)\,P^{\text{gt}}, \quad \boldsymbol{\eta}\sim\mathcal{N}(0,I),
\end{equation}
with the loss:
\begin{equation}
\mathcal{L}_3 = \mathbb{E}_{t,\boldsymbol{\eta},P^{\text{gt}}}\left[\|v_\theta^{(3)}(t,\mathbf{x}_t,\widetilde{X}_{\text{low}},\mathbf{z}_{\text{demo}},\text{pos})-(P^{\text{gt}}-\boldsymbol{\eta})\|^2\right].
\end{equation}

At inference, we synthesize the full high-resolution CT volume patch-by-patch, iteratively sampling overlapping regions. Each patch is independently generated starting from noise, conditioned appropriately, and then blended into the final volume.

\section{Experiments and Results}

\paragraph{Implementation}

In our implementation, we adopt the same neural network architecture for $v_\theta^{(1)}$ and $v_\theta^{(2)}$, while using a more compact model for $v_\theta^{(3)}$ due to computational constraints at super-resolution task. Specifically, each model is based on a 3D U-Net-like convolutional neural network modified from the EDM-img512-xxs design~\cite{karras2024analyzing}. Models $v_\theta^{(1)}$ and $v_\theta^{(2)}$ each contain approximately 80.68 million parameters, whereas the super-resolution model $v_\theta^{(3)}$ has approximately 1.89 million parameters by reducing the number of channels to 32 and the depth of 3D Unet.

Training was performed using AdamW optimization with an initial learning rate of $10^{-4}$, gradually reduced via linear decay. Specifically, we trained $v_\theta^{(1)}$ and $v_\theta^{(2)}$ each for approximately 35 million steps on four NVIDIA A100 GPUs (40GB each), requiring roughly 72 hours per stage. The high-resolution stage ($v_\theta^{(3)}$) was trained for approximately 15 million steps on the same hardware setup. The total batch size was 2048, and 8 samples were assigned per GPU with gradient accumulation. To enhance stability and convergence during training, we employed gradient clipping and maintained an exponential moving average (EMA) of the model weights as suggested from the original implementation of EDM2 \cite{karras2024analyzing}. Notably, the model pre-conditioning was not used for compatibility to flow-matching models. The flow matching objective utilized a standard $L_2$ regression loss between predicted and ground-truth velocity fields with a conditional optimal transport probability path \cite{lipman2024flowmatchingguidecode}.

For sampling, we discretized the temporal domain into 200 steps for all stages, transitioning from Gaussian noise ($\sigma_{\text{max}}=80$) using $\rho=7$. We employed the Dormand-Prince numerical solver with absolute and relative tolerances set to $\text{atol}=10^{-5}$ and $\text{rtol}=10^{-5}$, respectively. Computationally, models $v_\theta^{(1)}$ and $v_\theta^{(2)}$ required approximately 1319.92 Gflops per step and 0.59 GB GPU memory, whereas the super-resolution model $v_\theta^{(3)}$ used 183.10 Gflop sper step  and 0.20 GB GPU memory. All experiments were performed on eight NVIDIA A100 GPUs.

\begin{table}[ht]
\caption{Demographic and anthropometric characteristics of the study cohorts stratified by sex.}
\label{study_cohort}
\centering
\begin{tabular}{llrrrr}
\toprule
Cohort   & Sex    & $n$   & Age (yr)    & Height (cm) & Weight (kg) \\
\midrule
Training & Male   & 2950  & $69 \pm 10$ & $175 \pm 9$ & $90 \pm 17$ \\
         & Female &  248  & $58 \pm 17$ & $164 \pm 7$ & $71 \pm 16$ \\
Testing  & Male   &  608  & $67 \pm 11$ & $176 \pm 7$ & $89 \pm 17$ \\
         & Female &   92  & $59 \pm 15$ & $165 \pm 7$ & $76 \pm 20$ \\
\bottomrule
\end{tabular}

\end{table}

\paragraph{Dataset and Preprocessing}
Our models were trained and evaluated on a combined dataset comprising 3,198 torso 3D CT scans ($\simeq$ 1.13 million 2D axial slices), including anonymized 2,633 patients from Massachusetts General Hospital (MGH) and 565 from the publicly available AutoPET challenge dataset \cite{gatidis2022whole}. For evaluation, we reserved separate sets of 500 scans from MGH and 200 from AutoPET. The MGH dataset primarily consists of abdominal CT scans acquired from a male cancer patient cohort, while the AutoPET dataset comprises low-dose whole-body PET/CT scans collected from multiple institutions. This retrospective study was approved by the Institutional Review Board of Massachusetts General Hospital (Protocol 2022P001512), and the requirement for informed consent was waived. Detailed demographic information is provided in Table~\ref{study_cohort}.

We extracted a standardized torso region (spanning from the shoulders to mid-thigh) using TotalSegmentator~\cite{wasserthal2023totalsegmentator} from each CT and resampled all images to a uniform orientation and isotropic voxel spacing of $2\,\text{mm}$. To standardize intensity values, CTs were clipped in the range \([-500, 500]\) HU and linearly scaled to \([0, 1]\). To ensure consistent anatomical coverage, including key organs such as the heart, liver, and kidneys, CT volumes were segmented using the TotalSegmentator, and images were cropped to the region extending from the clavicle to the sacrum bones. During evaluation, normalized images were inverted to the original scale. Demographics were encoded as continuous values (age, height, weight), with sex encoded as a binary value.

To generate surface scans, we first segmented the external skin boundary resulting in a binary surface mask. We converted this mask into a triangular mesh and subsequently computed its signed distance function (SDF) representation on a discrete grid of size $56\times56\times88$ voxels. To simulate clinically realistic scenarios of partial surface acquisition—such as those encountered with single-view depth sensors—we introduced systematic incompleteness into the input surface meshes. For each subject, we identified the median axial slice of the torso and removed the entire posterior half of the surface mesh, thereby mimicking a frontal-only capture condition. 

\begin{figure} [b]
    \centering
    \includegraphics[width=1.0\textwidth]{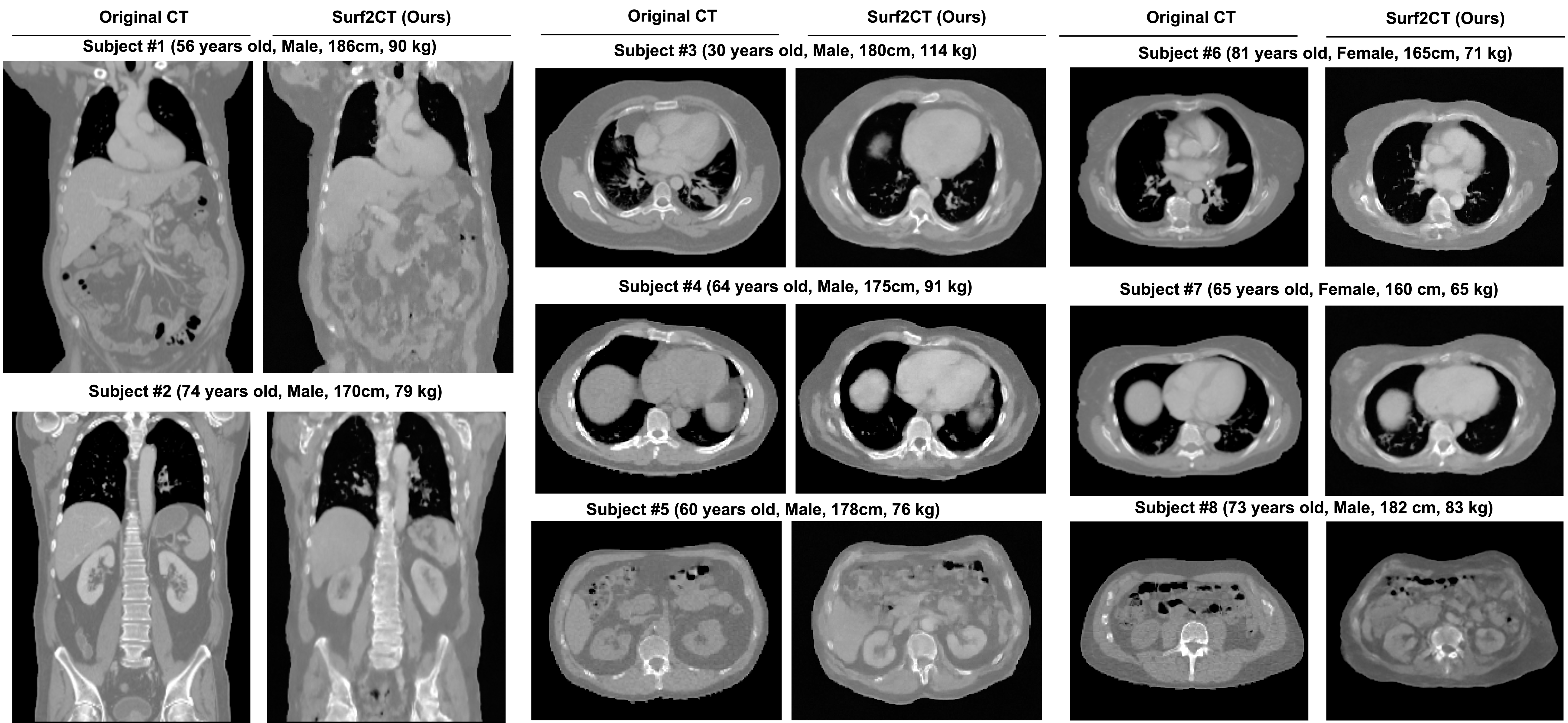}
    \caption{Qualitative comparison between original CT scans and Surf2CT-generated 3D volumes.}
    \label{fig:surf2ct_results}
\end{figure}
\section{Evaluation Setting and Results}

To evaluate the Surf2CT framework, we prioritize assessing anatomical realism and clinical plausibility rather than voxel-level fidelity. We therefore employ metrics reflecting anatomical consistency and clinical relevance. Our evaluation covers four potential applications:  body composition estimation, organ volume accuracy, and correlation, lung localization precision, and surface completion quality.

Anatomical structures for quantitative assessments were segmented using TotalSegmentator \cite{wasserthal2023totalsegmentator} for organ volumes and nnUnet2 \cite{chen2025automated} for body composition metrics. The reported Diff. (\%) is defined as Value of  (Surf2CT - Original CT) / Original CT × 100\%. We performed paired t-tests to assess statistical significance, using a threshold of 0.05. Results are reported as mean ± standard deviation, along with linear regression slopes and Pearson correlation coefficients. All evaluations were conducted using paired data consisting of the partial input surface and its corresponding original CT volume, enabling direct comparison and assessment of reconstruction accuracy under realistically incomplete input conditions.

\paragraph{Qualitative Assessment:}
Qualitative evaluation of Surf2CT-generated CT scans compared to corresponding original CT scans for multiple subjects (Fig. \ref{fig:surf2ct_results}). Coronal (subjects 1 and 2) and axial (subjects 3–8) views illustrate anatomical fidelity and consistency of synthesized images. Demographic data (age, sex, height, weight) provided for each subject demonstrates conditional modeling effectiveness in diverse patient characteristics.

\paragraph{Body Composition Evaluation:}

\begin{table} 
    \centering
    \caption{Volume of Body Composition Comparison (mL)}
    \label{tab:fat_comparison_transposed}
    \begin{tabular}{lccccc}
        \toprule
        & Original & Surf2CT & Diff. (\%) & Slope ($R^2$) \\
        \midrule
        \textbf{Male} & & & \\
        \quad Muscle & 9388 ± 1860 & 8483 ± 1424$^*$ & -9.6 & 0.69 (0.81) \\
        \quad Subcutaneous fat & 8288 ± 4189 & 9357 ± 3992$^*$ & +12.9 & 0.88 (0.86) \\
        \quad Visceral fat & 5030 ± 2472 & 4901 ± 2406 & -2.5 &0.84 (0.74)\\

        \midrule
        \textbf{Female} & & & \\
        \quad Muscle & 5304 ± 1152 & 4817 ± 993$^*$ & -9.2 & 0.80 (0.87)\\
        \quad Subcutaneous fat & 9210 ± 5368 & 9866 ± 4999 & +7.1 &  0.91 (0.96)\\
        \quad Visceral fat & 2324 ± 1514 & 2213 ± 1476 & -4.8 &0.87 (0.79)  \\
        \bottomrule
    \end{tabular}
\end{table}

\begin{figure}
    \centering
    \includegraphics[width=1.0\linewidth]{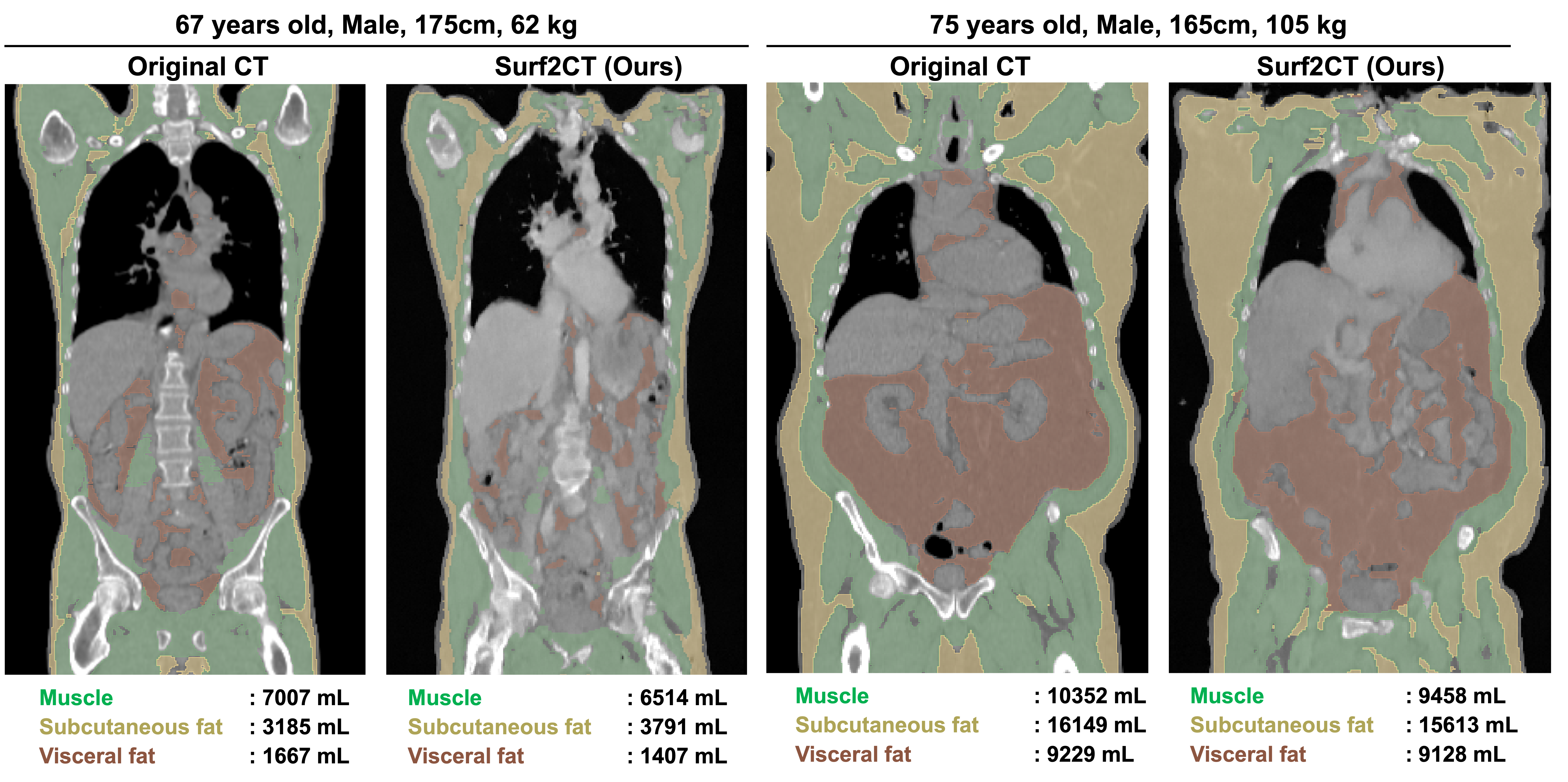}
    \caption{Body composition for two subjects comparing original CT and Surf2CT, highlighting consistent muscle, subcutaneous and visceral fat distributions across varying body types. }
    
    \label{fig:Fat_results}
\end{figure}

Two representative subjects’ body composition illustrated in Figure~\ref{fig:Fat_results} demonstrates qualitative agreement between Surf2CT-generated images and original CT scans across distinct body types. Quantitative evaluation (Table~\ref{tab:fat_comparison_transposed}, Appendix ~\ref{AppendixA}) shows that Surf2CT slightly underestimates skeletal muscle volume by approximately 9\% and modestly overestimates subcutaneous fat, with a greater deviation observed in males. Visceral fat estimates remain closely aligned with ground truth, differing by less than 5\%. Strong linear correlations across all body composition metrics, particularly in female subjects, highlight the predictive accuracy of Surf2CT (Table ~\ref{tab:fat_comparison_transposed}, Appendix ~\ref{AppendixA}). Additionally, demographic factors such as differences in muscle mass and fat mass by sex were observed. These findings indicate that Surf2CT can reproduce clinically relevant body composition profiles, supporting its potential utility in routine assessment.

\paragraph{Organ-wise Volume Evaluation:}

Surf2CT-generated organ volumes closely approximate ground-truth CT measurements across multiple organs and sexes (Table~\ref{tab:organ_comparison_transposed}, Appendix ~\ref{AppendixB}). As with the amount of muscle and fat, demographic factors were observed similarly to the original CT, with significant differences in the size of each organ by sex. Mean volume differences remained within ±5\% for most organs, including the liver and heart, with slightly larger deviations observed in the lungs and kidneys—particularly in female kidney volumes (-11.1\%). Statistical analysis revealed significant differences in male heart (\textit{p} < 0.001), male liver (\textit{p} < 0.001), total kidney (\textit{p} < 0.05), and male lung volumes (\textit{p} < 0.001). In contrast, organ volume differences in females were not statistically significant for the heart (\textit{p} = 0.239), liver (\textit{p} = 0.512), or lungs (\textit{p} = 0.079).

Despite the modest percentage errors, linear regression slopes and \( R^2 \) values were relatively low across organs (Table~\ref{tab:organ_comparison_transposed}, Appendix~\ref{AppendixB}), suggesting variability in subject-specific prediction accuracy. This may stem from anatomical heterogeneity in the training data, especially for internal organs not directly observable from surface geometry. These findings demonstrate that Surf2CT can generate anatomically plausible organ volumes from partial surface data, with better performance in female subjects and relatively small average volume errors across most organs.

\begin{table} 
    \centering
    \caption{Organ-wise volume comparison (mL)}
    \label{tab:organ_comparison_transposed}
    \begin{tabular}{lcccc}
        \toprule
        & Original & Surf2CT & Diff. (\%) & Slope ($R^2$) \\
        \midrule
        \textbf{Male} & & & \\
        \quad Heart & 689.7 ± 133.2 & 720.0 ± 84.6$^*$ & +4.4 & 0.23 (0.12) \\
        \quad Liver & 1695.4 ± 380.8 & 1761.1 ± 268.7$^*$ & +3.9 & 0.37 (0.25) \\
        \quad Kidney & 337.4 ± 70.1 & 329.1 ± 56.9$^*$ & -2.5 & 0.32 (0.16)\\
        \quad Lung & 3490.0 ± 727.2 & 3193.7 ± 439.8$^*$ & -8.5& 0.15 (0.06) \\
        \midrule
        \textbf{Female} & & & \\
        \quad Heart & 543.9 ± 82.9 & 557.0 ± 92.0 & +2.4 &0.36 (0.11) \\
        \quad Liver & 1524.9 ± 346.7 & 1495.7 ± 239.0 & -1.9 & 0.42 (0.33)\\
        \quad Kidney & 293.1 ± 55.3 & 260.5 ± 58.9$^*$ & -11.1 &0.18 (0.04) \\
        \quad Lung & 2532.1 ± 460.8 & 2431.4 ± 408.8 & -4.0 & 0.07 (0.01)\\
        \bottomrule
    \end{tabular}
\end{table}

\paragraph{Surface Completion Evaluation:}

\begin{figure}
    \centering
    \includegraphics[width=1.0\linewidth]{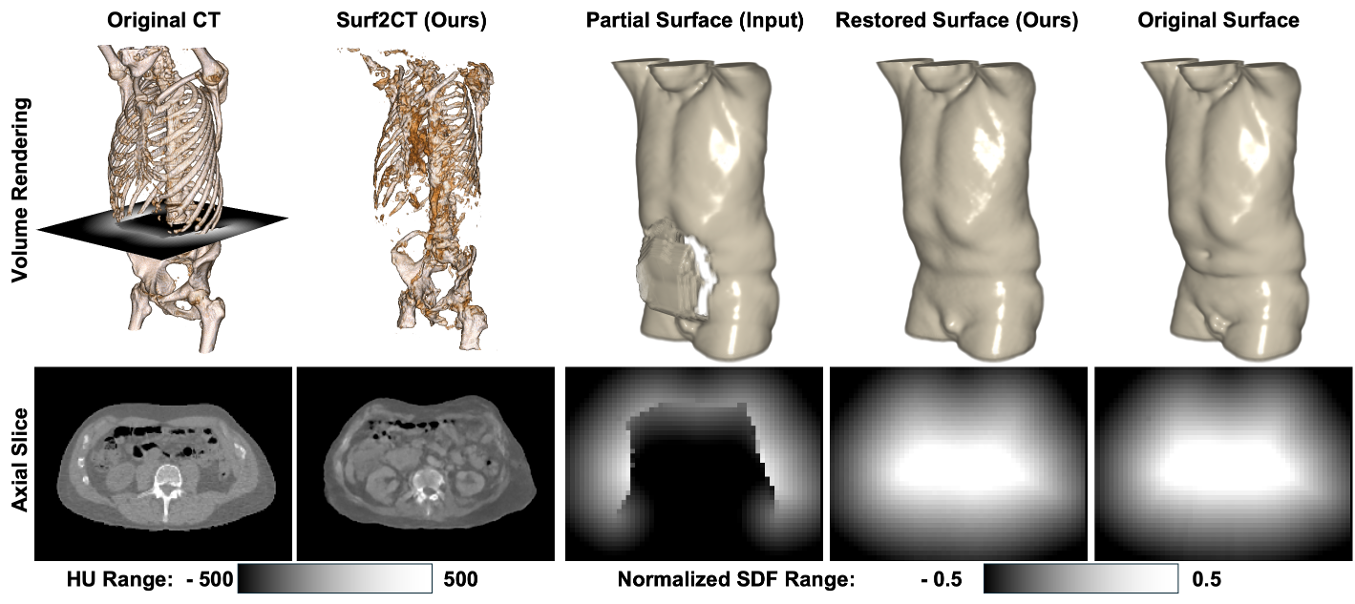}
    \caption{Qualitative visualization of surface completion. Top row illustrates volume-rendered CT alongside surface reconstruction results for partial input, restored, and original surfaces. Bottom row shows corresponding axial slices of normalized Signed Distance Functions (SDFs).}
    
    \label{fig:surface_completion}
\end{figure}

The performance of our surface completion method is visually illustrated in Figure~\ref{fig:surface_completion}. Our method significantly enhanced the quality of the reconstructed surfaces compared to the incomplete input (partial SDF). Specifically, the restored surfaces exhibited substantially lower Chamfer Distance (CD: $2.71 \pm 1.80$ vs. $521.78 \pm 228.09$), higher Intersection-over-Union (IoU: $0.98 \pm 0.02$ vs. $0.87 \pm 0.09$), and reduced Normalized Mean Absolute Error (NMAE: $0.02 \pm 0.01$ vs. $0.14 \pm 0.07$). These quantitative improvements highlight the effectiveness of our method in precise surface reconstructions from incomplete surface inputs.

\begin{figure}
    \centering
    \includegraphics[width=1\linewidth]{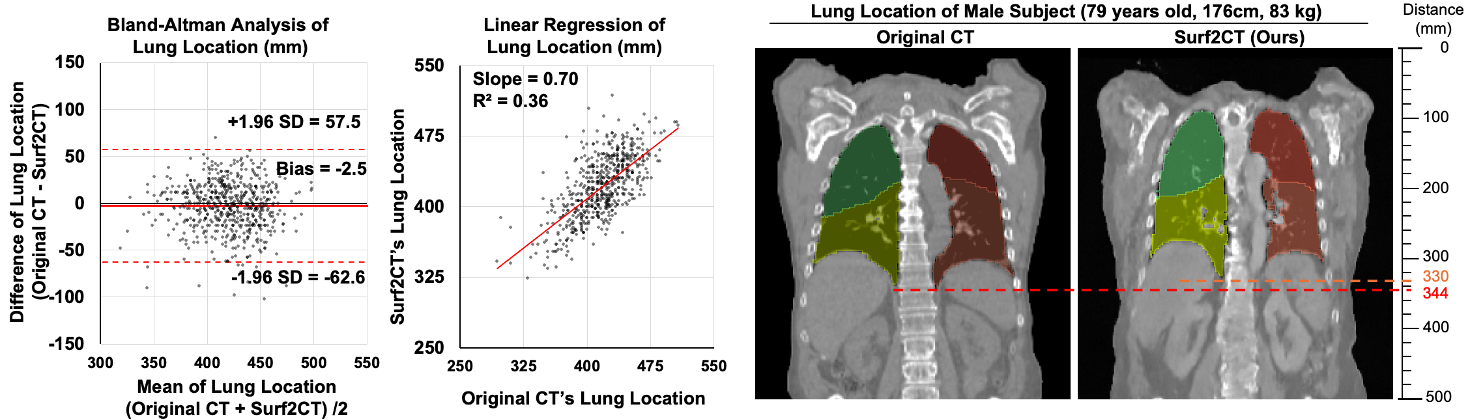}
    \caption{Quantitative (Bland-Altman, regression) and qualitative evaluation demonstrating accurate lung localization. The right panel visually compares lung localization in a representative subject with marked reference points indicating lung boundary positions.}
    
    \label{fig:lung_localization}
\end{figure}

\paragraph{Lung Localization Evaluation:} 
Lung localization was quantified as the vertical distance between the shoulder apex and the lowest extent of the segmented lung, derived using TotalSegmentator, and is both numerically evaluated and visually shown in Figure~\ref{fig:lung_localization}. Bland-Altman analysis revealed minimal bias (-2.5 mm) in lung localization measurements between original CT and Surf2CT-generated images, with limits of agreement ranging from -62.6 mm to +57.5 mm. Linear regression analysis further confirmed a moderate correlation (slope = 0.70, $R^2$ = 0.36). The representative subject visualization demonstrates good spatial alignment of lung anatomy in Surf2CT-generated images compared to original CT, illustrating Surf2CT’s potential to support reliable anatomical localization. 

\section{Discussion}

We introduced Surf2CT, a generative model designed to accurately synthesize clinically relevant 3D CT volumes from external surface scans combined with demographic information. Surf2CT demonstrated robust anatomical realism and clinical plausibility across multiple quantitative evaluations, including organ volume estimation, body composition analysis, surface completion accuracy, and precise lung localization. Our results showed that Surf2CT leverages external body surface and basic demographics to generate detailed internal anatomical representations. This capability significantly expands the potential for non-invasive anatomical assessments, positioning Surf2CT as a promising tool for routine clinical screening, personalized health monitoring, and preventive medical practices.

Nevertheless, the current Surf2CT framework has limitations due to its training on predominantly cancer patient's CT datasets. This constraint inherently restricts the model’s capability to represent various pathologies that minimally alter external morphology, thus limiting its diagnostic applicability in clinical settings where internal anomalies are suspected. However, interestingly, deviations between synthetic predictions and actual patient anatomy could potentially flag hidden abnormalities. Future studies could investigate this diagnostic adjunct potential, particularly in scenarios where substantial discrepancies in predicted versus actual organ volumes or body composition occur.

Another important consideration involves the discrepancy between training and deployment conditions. Real-world surface scanning methods, such as consumer-grade scanners (e.g., Kinect, smartphone-based scanners), inherently introduce varying degrees of noise and artifacts not represented in the current training data.  Future work should explore the integration of additional informative inputs, such as disease or text data, to further personalize to other modalities, such as PET and MRI, and refine predictions. It will also be essential to address practical deployment challenges, including robust handling of real-world scanner variability, and ensuring the generalizability and fairness of the model across diverse patient populations. Surf2CT exemplifies how innovative generative modeling approaches can bridge external physiological observations with comprehensive internal health assessments.

\bibliographystyle{unsrtnat}  % unsorted numeric with natbib
\bibliography{Paper}

\begin{thebibliography}{26}
\providecommand{\natexlab}[1]{#1}
\providecommand{\url}[1]{\texttt{#1}}
\expandafter\ifx\csname urlstyle\endcsname\relax
  \providecommand{\doi}[1]{doi: #1}\else
  \providecommand{\doi}{doi: \begingroup \urlstyle{rm}\Url}\fi

\bibitem[Smith-Bindman et~al.(2009)Smith-Bindman, Lipson, Marcus, Kim, Mahesh, Gould, De~Gonz{\'a}lez, and Miglioretti]{smith2009radiation}
Rebecca Smith-Bindman, Jafi Lipson, Ralph Marcus, Kwang-Pyo Kim, Mahadevappa Mahesh, Robert Gould, Amy~Berrington De~Gonz{\'a}lez, and Diana~L Miglioretti.
\newblock Radiation dose associated with common computed tomography examinations and the associated lifetime attributable risk of cancer.
\newblock \emph{Archives of internal medicine}, 169\penalty0 (22):\penalty0 2078--2086, 2009.

\bibitem[Treleaven and Wells(2007)]{treleaven20073d}
Philip Treleaven and Jonathan Wells.
\newblock 3d body scanning and healthcare applications.
\newblock \emph{Computer}, 40\penalty0 (7):\penalty0 28--34, 2007.

\bibitem[Lee et~al.(2021)Lee, Hong, Witanto, Choi, Park, Decazes, Eude, Kim, Kim, Goo, et~al.]{lee2021deep}
Yoon~Seong Lee, Namki Hong, Joseph~Nathanael Witanto, Ye~Ra Choi, Junghoan Park, Pierre Decazes, Florian Eude, Chang~Oh Kim, Hyeon~Chang Kim, Jin~Mo Goo, et~al.
\newblock Deep neural network for automatic volumetric segmentation of whole-body ct images for body composition assessment.
\newblock \emph{Clinical Nutrition}, 40\penalty0 (8):\penalty0 5038--5046, 2021.

\bibitem[Despr{\'e}s et~al.(2008)Despr{\'e}s, Lemieux, Bergeron, Pibarot, Mathieu, Larose, Rod{\'e}s-Cabau, Bertrand, and Poirier]{despres2008abdominal}
Jean-Pierre Despr{\'e}s, Isabelle Lemieux, Jean Bergeron, Philippe Pibarot, Patrick Mathieu, Eric Larose, Josep Rod{\'e}s-Cabau, Olivier~F Bertrand, and Paul Poirier.
\newblock Abdominal obesity and the metabolic syndrome: contribution to global cardiometabolic risk.
\newblock \emph{Arteriosclerosis, thrombosis, and vascular biology}, 28\penalty0 (6):\penalty0 1039--1049, 2008.

\bibitem[Cruz-Jentoft and Sayer(2019)]{cruz2019sarcopenia}
Alfonso~J Cruz-Jentoft and Avan~A Sayer.
\newblock Sarcopenia.
\newblock \emph{The Lancet}, 393\penalty0 (10191):\penalty0 2636--2646, 2019.

\bibitem[Asciak et~al.(2025)Asciak, Kyeremeh, Luo, Kazakidi, Connolly, Picard, O’Neill, Tsaftaris, Stewart, and Shu]{asciak2025digital}
Lisa Asciak, Justicia Kyeremeh, Xichun Luo, Asimina Kazakidi, Patricia Connolly, Frederic Picard, Kevin O’Neill, Sotirios~A Tsaftaris, Grant~D Stewart, and Wenmiao Shu.
\newblock Digital twin assisted surgery, concept, opportunities, and challenges.
\newblock \emph{npj Digital Medicine}, 8\penalty0 (1):\penalty0 32, 2025.

\bibitem[Song and Ermon(2019)]{song2019generative}
Yang Song and Stefano Ermon.
\newblock Generative modeling by estimating gradients of the data distribution.
\newblock \emph{Advances in neural information processing systems}, 32, 2019.

\bibitem[Lipman et~al.(2022)Lipman, Chen, Ben-Hamu, Nickel, and Le]{lipman2022flow}
Yaron Lipman, Ricky~TQ Chen, Heli Ben-Hamu, Maximilian Nickel, and Matt Le.
\newblock Flow matching for generative modeling.
\newblock \emph{arXiv preprint arXiv:2210.02747}, 2022.

\bibitem[Vahdat et~al.(2022)Vahdat, Williams, Gojcic, Litany, Fidler, Kreis, et~al.]{vahdat2022lion}
Arash Vahdat, Francis Williams, Zan Gojcic, Or~Litany, Sanja Fidler, Karsten Kreis, et~al.
\newblock Lion: Latent point diffusion models for 3d shape generation.
\newblock \emph{Advances in Neural Information Processing Systems}, 35:\penalty0 10021--10039, 2022.

\bibitem[Galvis et~al.(2024)Galvis, Zuo, Schaefer, and Leutengger]{galvis2024sc}
Juan~D Galvis, Xingxing Zuo, Simon Schaefer, and Stefan Leutengger.
\newblock Sc-diff: 3d shape completion with latent diffusion models.
\newblock \emph{arXiv preprint arXiv:2403.12470}, 2024.

\bibitem[Guo et~al.(2025)Guo, Zhao, Yang, Xu, Nath, Tang, Simon, Belue, Harmon, Turkbey, et~al.]{guo2025maisi}
Pengfei Guo, Can Zhao, Dong Yang, Ziyue Xu, Vishwesh Nath, Yucheng Tang, Benjamin Simon, Mason Belue, Stephanie Harmon, Baris Turkbey, et~al.
\newblock Maisi: Medical ai for synthetic imaging.
\newblock In \emph{2025 IEEE/CVF Winter Conference on Applications of Computer Vision (WACV)}, pages 4430--4441. IEEE, 2025.

\bibitem[Shen et~al.(2019)Shen, Zhao, and Xing]{shen2019patient}
Liyue Shen, Wei Zhao, and Lei Xing.
\newblock Patient-specific reconstruction of volumetric computed tomography images from a single projection view via deep learning.
\newblock \emph{Nature biomedical engineering}, 3\penalty0 (11):\penalty0 880--888, 2019.

\bibitem[Pan et~al.(2024)Pan, Abouei, Wynne, Chang, Wang, Qiu, Li, Peng, Roper, Patel, et~al.]{pan2024synthetic}
Shaoyan Pan, Elham Abouei, Jacob Wynne, Chih-Wei Chang, Tonghe Wang, Richard~LJ Qiu, Yuheng Li, Junbo Peng, Justin Roper, Pretesh Patel, et~al.
\newblock Synthetic ct generation from mri using 3d transformer-based denoising diffusion model.
\newblock \emph{Medical Physics}, 51\penalty0 (4):\penalty0 2538--2548, 2024.

\bibitem[Segars et~al.(2010)Segars, Sturgeon, Mendonca, Grimes, and Tsui]{segars20104d}
W~Paul Segars, G~Sturgeon, S~Mendonca, Jason Grimes, and Benjamin~MW Tsui.
\newblock 4d xcat phantom for multimodality imaging research.
\newblock \emph{Medical physics}, 37\penalty0 (9):\penalty0 4902--4915, 2010.

\bibitem[Shetty et~al.(2023)Shetty, Birkhold, Jaganathan, Strobel, Egger, Kowarschik, and Maier]{shetty2023boss}
Karthik Shetty, Annette Birkhold, Srikrishna Jaganathan, Norbert Strobel, Bernhard Egger, Markus Kowarschik, and Andreas Maier.
\newblock Boss: Bones, organs and skin shape model.
\newblock \emph{Computers in Biology and Medicine}, 165:\penalty0 107383, 2023.

\bibitem[Ying et~al.(2019)Ying, Guo, Ma, Wu, Weng, and Zheng]{ying2019x2ct}
Xingde Ying, Heng Guo, Kai Ma, Jian Wu, Zhengxin Weng, and Yefeng Zheng.
\newblock X2ct-gan: reconstructing ct from biplanar x-rays with generative adversarial networks.
\newblock In \emph{Proceedings of the IEEE/CVF conference on computer vision and pattern recognition}, pages 10619--10628, 2019.

\bibitem[Sun et~al.(2024)Sun, Baroudi, Netherton, Court, Mawlawi, Veeraraghavan, and Balakrishnan]{sun2024difr3ct}
Yiran Sun, Hana Baroudi, Tucker Netherton, Laurence Court, Osama Mawlawi, Ashok Veeraraghavan, and Guha Balakrishnan.
\newblock Difr3ct: Latent diffusion for probabilistic 3d ct reconstruction from few planar x-rays.
\newblock \emph{arXiv preprint arXiv:2408.15118}, 2024.

\bibitem[Shen et~al.(2022)Shen, Zhao, Capaldi, Pauly, and Xing]{shen2022geometry}
Liyue Shen, Wei Zhao, Dante Capaldi, John Pauly, and Lei Xing.
\newblock A geometry-informed deep learning framework for ultra-sparse 3d tomographic image reconstruction.
\newblock \emph{Computers in Biology and Medicine}, 148:\penalty0 105710, 2022.

\bibitem[Yang et~al.(2025)Yang, Chen, Sun, Strittmatter, Raj, Allababidi, Rink, and Z{\"o}llner]{yang2025seg2med}
Zeyu Yang, Zhilin Chen, Yipeng Sun, Anika Strittmatter, Anish Raj, Ahmad Allababidi, Johann~S Rink, and Frank~G Z{\"o}llner.
\newblock seg2med: a segmentation-based medical image generation framework using denoising diffusion probabilistic models.
\newblock \emph{arXiv preprint arXiv:2504.09182}, 2025.

\bibitem[Chu et~al.(2023)Chu, Xie, Mo, Li, Nie{\ss}ner, Fu, and Jia]{chu2023diffcomplete}
Ruihang Chu, Enze Xie, Shentong Mo, Zhenguo Li, Matthias Nie{\ss}ner, Chi-Wing Fu, and Jiaya Jia.
\newblock Diffcomplete: Diffusion-based generative 3d shape completion.
\newblock \emph{Advances in neural information processing systems}, 36:\penalty0 75951--75966, 2023.

\bibitem[Wang et~al.(2023)Wang, Jiang, Zheng, Wang, He, Wang, Chen, Zhou, et~al.]{wang2023patch}
Zhendong Wang, Yifan Jiang, Huangjie Zheng, Peihao Wang, Pengcheng He, Zhangyang Wang, Weizhu Chen, Mingyuan Zhou, et~al.
\newblock Patch diffusion: Faster and more data-efficient training of diffusion models.
\newblock \emph{Advances in neural information processing systems}, 36:\penalty0 72137--72154, 2023.

\bibitem[Karras et~al.(2024)Karras, Aittala, Lehtinen, Hellsten, Aila, and Laine]{karras2024analyzing}
Tero Karras, Miika Aittala, Jaakko Lehtinen, Janne Hellsten, Timo Aila, and Samuli Laine.
\newblock Analyzing and improving the training dynamics of diffusion models.
\newblock In \emph{Proceedings of the IEEE/CVF Conference on Computer Vision and Pattern Recognition}, pages 24174--24184, 2024.

\bibitem[Lipman et~al.(2024)Lipman, Havasi, Holderrieth, Shaul, Le, Karrer, Chen, Lopez-Paz, Ben-Hamu, and Gat]{lipman2024flowmatchingguidecode}
Yaron Lipman, Marton Havasi, Peter Holderrieth, Neta Shaul, Matt Le, Brian Karrer, Ricky T.~Q. Chen, David Lopez-Paz, Heli Ben-Hamu, and Itai Gat.
\newblock Flow matching guide and code, 2024.
\newblock URL \url{https://arxiv.org/abs/2412.06264}.

\bibitem[Gatidis et~al.(2022)Gatidis, Hepp, Fr{\"u}h, La~Foug{\`e}re, Nikolaou, Pfannenberg, Sch{\"o}lkopf, K{\"u}stner, Cyran, and Rubin]{gatidis2022whole}
Sergios Gatidis, Tobias Hepp, Marcel Fr{\"u}h, Christian La~Foug{\`e}re, Konstantin Nikolaou, Christina Pfannenberg, Bernhard Sch{\"o}lkopf, Thomas K{\"u}stner, Clemens Cyran, and Daniel Rubin.
\newblock A whole-body fdg-pet/ct dataset with manually annotated tumor lesions.
\newblock \emph{Scientific Data}, 9\penalty0 (1):\penalty0 601, 2022.

\bibitem[Wasserthal et~al.(2023)Wasserthal, Breit, Meyer, Pradella, Hinck, Sauter, Heye, Boll, Cyriac, Yang, et~al.]{wasserthal2023totalsegmentator}
Jakob Wasserthal, Hanns-Christian Breit, Manfred~T Meyer, Maurice Pradella, Daniel Hinck, Alexander~W Sauter, Tobias Heye, Daniel~T Boll, Joshy Cyriac, Shan Yang, et~al.
\newblock Totalsegmentator: robust segmentation of 104 anatomic structures in ct images.
\newblock \emph{Radiology: Artificial Intelligence}, 5\penalty0 (5):\penalty0 e230024, 2023.

\bibitem[Chen et~al.(2025)Chen, Gu, Chen, Yang, Dong, Cao, Camarena, Mantyh, Colglazier, and Mazurowski]{chen2025automated}
Yaqian Chen, Hanxue Gu, Yuwen Chen, Jicheng Yang, Haoyu Dong, Joseph~Y Cao, Adrian Camarena, Christopher Mantyh, Roy Colglazier, and Maciej~A Mazurowski.
\newblock Automated muscle and fat segmentation in computed tomography for comprehensive body composition analysis.
\newblock \emph{arXiv preprint arXiv:2502.09779}, 2025.

\end{thebibliography}

\medskip

%%%%%%%%%%%%%%%%%%%%%%%%%%%%%%%%%%%%%%%%%%%%%%%%%%%%%%%%%%%%

\appendix

\newpage

\section{Appendix A: Violin Plot and Linear Regression Analysis of Body Composition}
\label{AppendixA}

\begin{figure}[htbp]
\centering
    \includegraphics[width=1\linewidth]{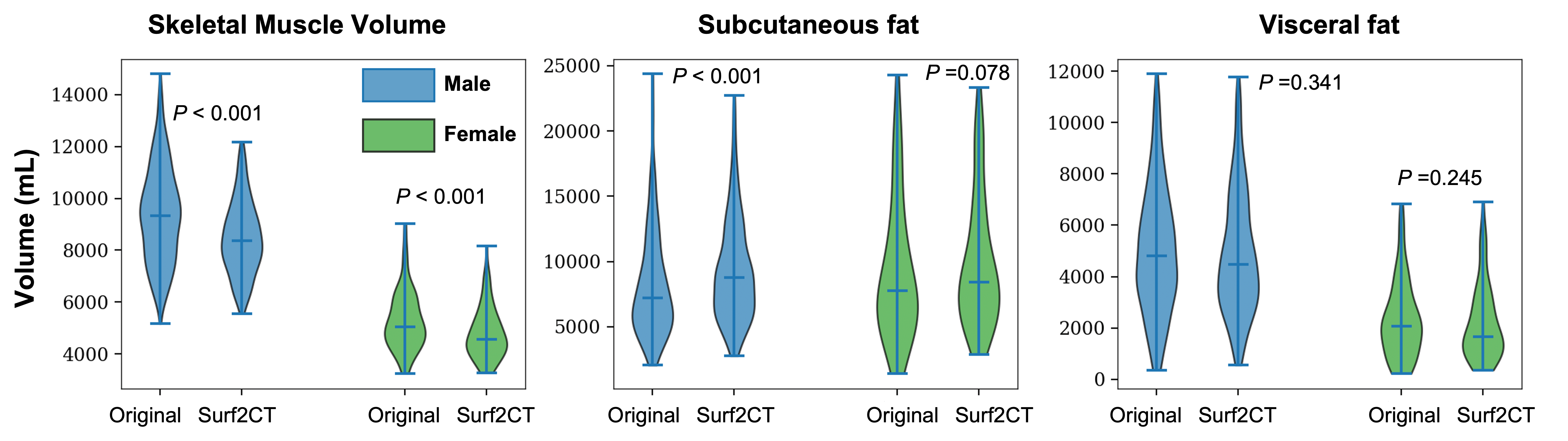}
    \caption{Violin plots showing comparisons between original and Surf2CT-derived volumes of skeletal muscle, subcutaneous fat, and visceral fat, separated by sex. }
    
    \label{fig:Fat_violin}
\end{figure}

\begin{figure}[htbp]
\centering
\includegraphics[width=0.7\linewidth]{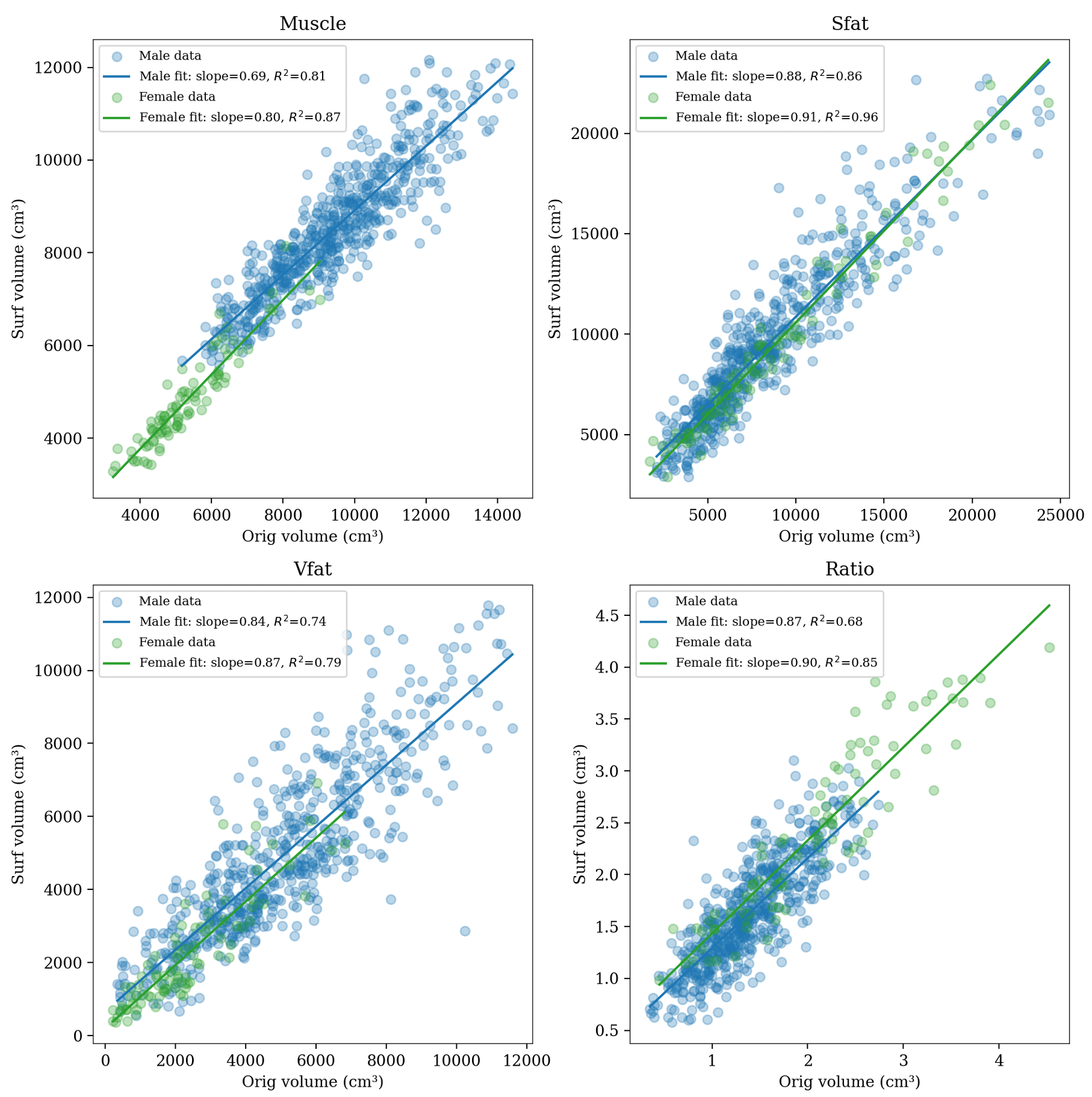}
\caption{Linear regression plots comparing Surf2CT-derived and original CT volumes for skeletal muscle, subcutaneous fat, visceral fat, and muscle-to-total fat ratio. Regression lines and corresponding slopes and $R^2$ values are indicated separately for male and female subjects.}
\label{fig:regression_body_composition}
\end{figure}

To quantitatively assess the agreement between Surf2CT-generated and original CT-derived body composition, violin Plot and linear regression analyses were performed.  Figures~\ref{fig:Fat_violin} illustrate Violin plots and Figures~\ref{fig:regression_body_composition} illustrate linear regression analyses.

\textbf{Body Composition Analysis:} As demonstrated in Figure~\ref{fig:Fat_violin} and Figure~\ref{fig:regression_body_composition}, the regression analysis revealed robust correlations across all assessed body composition metrics. Particularly high coefficients of determination ($R^2$) were observed for subcutaneous fat (male: 0.86, female: 0.96) and skeletal muscle volumes (male: 0.81, female: 0.87). Visceral fat showed slightly lower but still robust correlations (male: 0.74, female: 0.79), indicating consistent predictive capability by Surf2CT across different body types

\newpage
\section{Appendix B: Violin Plot and Linear Regression Analysis of Organ Volumes}

\label{AppendixB}

\begin{figure}[htbp]
    \centering
    \includegraphics[width=1\linewidth]{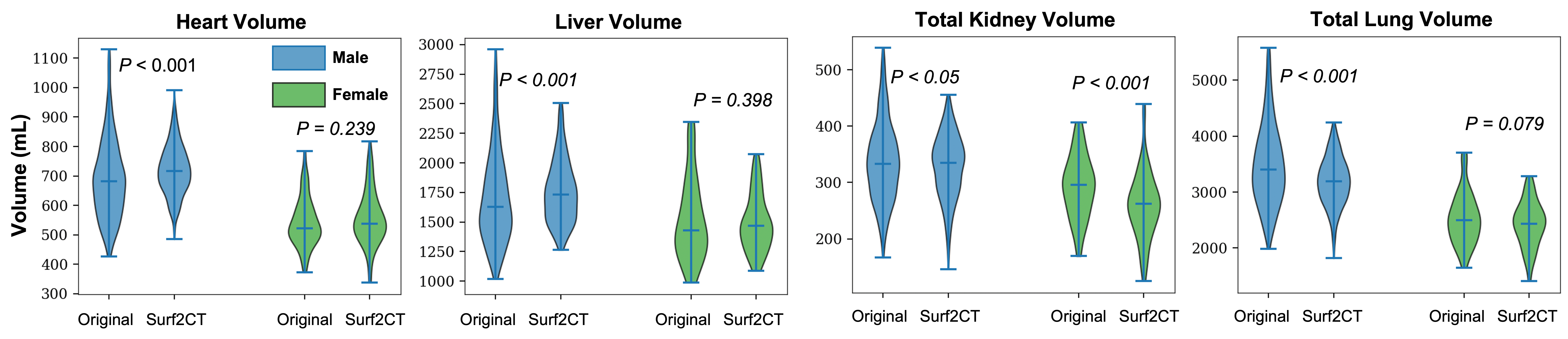}
    \caption{Comparison of organ volumes between original CT and Surf2CT stratified by sex. Violin plots show volume distributions for the heart, liver, kidney, and lung.}
    
    \label{fig:Organ_violin}
\end{figure}

\begin{figure}[htbp]
\centering
\includegraphics[width=0.7\linewidth]{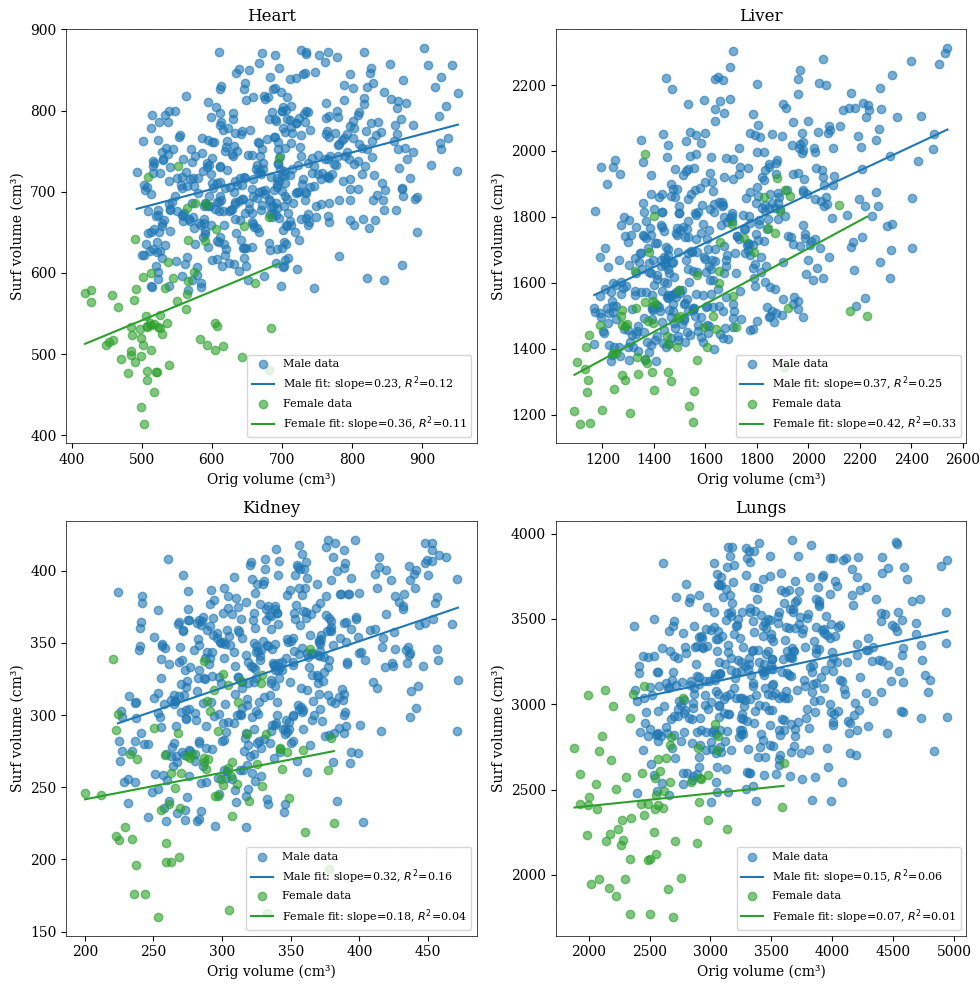}
\caption{Linear regression plots comparing Surf2CT-derived and original CT organ volumes for the heart, liver, kidney, and lungs. Regression lines and corresponding slopes and $R^2$ values are shown separately for male and female subjects.}
\label{fig:regression_organ_volume}
\end{figure}

To quantitatively assess the agreement between Surf2CT-generated and original CT-derived organ volumes, violin plot and linear regression analyses were performed. Figures~\ref{fig:Organ_violin} illustrate violin plots and Figures~\ref{fig:regression_organ_volume} illustrate linear regression analyses.

\textbf{Organ-wise Volume Analysis:} Figure~\ref{fig:regression_organ_volume} summarizes the regression analysis for various organs. Generally lower $R^2$ values were observed compared to body composition metrics, reflecting inherent anatomical variability among subjects. Heart volumes showed the weakest correlations (male: $R^2$=0.12, female: $R^2$=0.11), likely due to variability in heart chamber sizes and functional states at scan time. Liver and kidney volumes displayed modest correlations, with liver volume correlations (male: 0.25, female: 0.33) slightly outperforming kidney volumes (male: 0.16, female: 0.04). Lung volume correlations were notably low (male: 0.06, female: 0.01), reflecting the high inter-individual anatomical variability and potential physiological differences impacting lung expansion and aeration.

%%%%%%%%%%%%%%%%%%%%%%%%%%%%%%%%%%%%%%%%%%%%%%%%%%%%%%%%%%%%

\end{document}